\documentclass[twocolumn,showpacs,aps,pre,superscriptaddress,floatfix,longbibliography]{revtex4-2}

\usepackage{amsmath,amsfonts,amssymb,amsthm,amstext,mathtools}
\usepackage{graphics,graphicx,epsfig,bbm,epsfig,epstopdf,psfrag}
\usepackage{dcolumn,bm,comment,dsfont,times,dsfont,braket,euscript}
\usepackage{subfigure, setspace}

\usepackage[colorlinks=true,citecolor=blue,linkcolor=blue,urlcolor=blue]{hyperref}
\usepackage[usenames]{color}

\usepackage{wrapfig}
\usepackage{sidecap}

\UseRawInputEncoding

\usepackage{lineno}
\usepackage[nameinlink]{cleveref}
\crefname{equation}{Eq.}{}
\Crefname{equation}{Eqs.}{}
\crefname{figure}{Figs.}{Figs.}
\Crefname{figure}{Fig.}{Fig.}

\crefname{table}{Table}{Table}
\crefname{Appendix}{Appendix}{Appendices}

\begin{document}
\title{Confined floating active carpets generate coherent vortical flows that enhance transport}

\date{October 29, 2025}

\author{Felipe A. Barros}
\email{felipe.barrosca@gmail.com}
\affiliation{Departamento de F\'isica, Facultad de Ciencias, Universidad de Chile, Santiago ,Chile}

\author{Italo Salas}
\email{italo.salas@ug.uchile.cl}
\affiliation{Departamento de F\'isica, Facultad de Ciencias, Universidad de Chile, Santiago ,Chile}

\author{Enkeleida Lushi}
\email{lushi@softactivematterlab.com}
\affiliation{Soft Active Matter Lab, Branchburg, NJ, 08876, United States}

\author{Francisca Guzm\'an-Lastra}
\email{fguzman@uchile.cl}
\affiliation{Departamento de F\'isica, Facultad de Ciencias, Universidad de Chile, Santiago ,Chile}

\begin{abstract}

Slicks are thin viscous films that can be found at the air--water interface of water bodies such as lakes, rivers and oceans. These micro-layers are enriched in surfactants, organic matter, and microorganisms, and exhibit steep physical and chemical gradients across only tens to hundreds of micrometers. In such geometrically confined environments, the hydrodynamics and transport of nutrients, pollutants, and microorganisms are constrained, yet they collectively sustain key biogenic processes. It remains however largely unexplored how the hydrodynamic flows and transport are affected by the vertical extent of slicks relative to the size of microbial colonies. Here, we study this question by combining analytical and numerical approaches to model a microbial colony as an active carpet: a two-dimensional distribution of micro-swimmers exerting dipolar forces. We show that there exists a ratio between the carpet size and the  confinement height that is optimal for the enhancement of particle transport toward the colony edges through advective flows that recirculate in 3D vortex-ring-like patterns with a characteristic length comparable to the confinement height. Our results demonstrate that finite, coherent vortex-ring-like structures can arise solely from the geometrical confinement ratio of slick thickness to microbial colony size. These findings shed light on the interplay between collective activity and out-of-equilibrium transport, and on how microbial communities form, spread, and persist in geometrically constrained environments such as surface slicks.

\end{abstract}


\maketitle

\section{Introduction}

Large bodies of water such as oceans, lakes, and rivers are complex systems, whose environmental dynamics are fundamentally influenced by countless interactions at the microscale, including microbial life \cite{stocker2012marine, Brumley20}. The sea surface micro-layer (SML) acts as the main precursor for these interactions, as this 1 to 1000 microns thick fluid layer lies just at the ocean-atmosphere interface, playing a key role in biogeochemical processes \cite{mustaffa2018high}. A characteristic phenomenon of the SML is the formation of surface slicks: macroscopic patches with visible difference in color, texture, commonly oily, and more biochemically complex than the non-slick fluid nearby. Slicks, that can be of natural or anthropogenic origin, and range from 0.04 to 200 microns thick and up to 10 km in length, serve as habitats for special microbial communities that are a primary source of the energy transported through the whole water column \cite{wurl2016biofilm,voskuhl2022natural}. In fact, microorganisms can aggregate and form biofilms of a much larger scale than a single cell in response to environmental hostility, thereby maintaining functionality and surviving at fluid interfaces \cite{hall2004bacterial,lopez2010biofilms,niepa2017films}. These self-sustained floating active suspensions, known as \textit{active carpets}, overcome physically challenging environmental stresses--such as geometrical constraints and changes in temperature, salinity, pressure, and viscosity--through interactions with each other and their surroundings. Such interactions often drive these systems into unexpected collective behaviors, pushing them far from thermal equilibrium \cite{mathijssen2018nutrient,guzman2021active,aguayo2024floating,barros2025layered}. However, how microbial carpets act and adapt within confined viscous environments such as surface slicks remains unexplored.

Swimming microorganisms self-propel autonomously through media by exerting mechanical forces and generating hydrodynamic flow fields \cite{Pedley92,turner2000real,polin2009chlamydomonas,lauga2009hydrodynamics,guasto2010oscillatory,drescher2010direct,drescher2011fluid,Kiorboe14,gilpin2017vortex,Guzman-Lastra25}. Therefore, their individual and collective behavior is influenced by short-range steric and long-range hydrodynamic interactions, both between cells and at cell-interface contacts \cite{Sokolov09,rabani2013collective,lushi2014fluid,Spagnolie15,Sipos15,Wioland16, Lushi17,Spagnolie23}. For instance, \textit{E. coli} bacteria are known to accumulate and be attracted to both solid surfaces and gas and liquid interfaces due to those interactions and reorientation processes \cite{lauga2006swimming,berke2008hydrodynamic,lemelle2010counterclockwise,Mino11,vaccari2017films,ahmadzadegan2019hydrodynamic}. Similarly, recent experiments have confirmed that phototactic microalgae \textit{C. reindhartii} accumulate at solid-liquid interfaces \cite{buchner2021hopping,li2025hydrodynamic}. Emergent behaviors like self-organization (e.g. bacterial and algal turbulence) are characteristic of these microbial multicellular aggregates, having several effects on their environment \cite{cho2007self,copeland2009bacterial,vincenti2019magnetotactic,peng2021imaging,baruah2025emergent,hokmabad2025spatial}: they can drive large-scale fluid flows and directed material transport \cite{xu2019self,li2024robust}, enhance diffusive transport and control of cargo---from small nutrient molecules to entire cells and much larger particles---through collective entrainment \cite{Kim04,short2006flows,mino2011enhanced,patteson2016particle,jeanneret2016entrainment,Brumley19,jin2021collective,grossmann2024non,laroussi2025controlling}, generate colloidal (and cellular) aggregation and self-assembly \cite{gokhale2022dynamic,grober2023unconventional,kushwaha2023phase,espada2024active}, induce coherent bioconvective patterns and benefit biomixing \cite{pedley1988growth,Tuval05,Kim07,Ezhilan12,sommer2017bacteria,dervaux2017light,arrieta2019light,thery2020self,gore2025oxygen,zhu2025turbulent}.

A common feature of the hydrodynamic responses of these active fluids is that they arise under some form of physical confinement --- from emulsion droplets to controllable lab-made chambers, often aiming to emulate real, constrained and varying environmental conditions \cite{du2021bacterial,fan2021effects}. At a single-cell level, experiments have shown that the flow fields generated by strongly confined microalgae change and enhance nutrient flux to the microorganism despite higher friction with the boundaries \cite{mondal2021strong}. Active-passive mixtures of microalgae and colloids confined to a channel lead the latter to complex steady-state distributions in a de-mixing process \cite{williams2022confinement} and more recently it was shown that  confined bacterial suspensions spatially self-organize from a immotile inner core to a motile shell in order to regulate oxygen transport, strictly dependent on suspension density and size \cite{hokmabad2025spatial}. These responses are also shaped by the properties of the fluid medium, where viscosity plays a dominant role. This is especially true at the SML, where viscosity gradients emerge due to temperature changes, among other \cite{voskuhl2022natural}. Such gradients influence microbial motility and nutrient transport, while also triggering biological activity and microbial adaptations. For instance, experiments have shown that blooms of \textit{P. globosa}, a colonial marine plankton, can extend up to 1 km and increase the surrounding viscosity by more than an order of magnitude \cite{guadayol2021microrheology}. Moreover, microbes exploit viscotaxis to survive in viscous mucus layers: in coral reefs, where viscosity is more than twice that of seawater and contributes to nutrient redistribution, and in the mammalian gut and trachea, where mucus can reach up to three orders of magnitude higher than the surrounding fluid and serves as a defense against microbial pathogens \cite{lai2009micro,stabili2015mucus,liebchen2018viscotaxis,stehnach2021viscophobic,ishikawa2025physics}.

A slick-like environment represents a less explored configuration that encapsulates many of these physical stimuli: a confined interfacial microbial layer characterized by viscous heterogeneity. Some theoretical studies have begun to address these \textit{floating} biofilms. Desai et al.~\cite{desai2020biofilms} developed a hydrodynamic model to investigate microbial distributions as functions of morphology, propulsion mechanism, and viscous heterogeneity, providing insights into swimming stability and revealing the preferential accumulation of elongated micro-swimmers at liquid--liquid interfaces \cite{aderogba1978action,desai2020biofilms}. More recently, Barros et al.~\cite{barros2025layered} theoretically examined the influence of an extended active carpet of micro-swimmers on the boundary conditions of floating biofilms. Their study explored spatial non-equilibrium fluctuations and transport as functions of confinement size and viscous heterogeneity, revealing anisotropic fluctuation distributions \cite{mathijssen2018nutrient,barros2025layered}. These contributions have expanded our understanding of microbial distributions and long-range collective fluctuations in floating biofilms. However, several questions remain unresolved: What types of collective flows emerge within these layers? Do they depend on the interplay between aggregate size and confinement scale? And can non-equilibrium transport driven by active carpet flows under confinement be described independently of length scales?

Here we investigate the emergence of three-dimensional hydrodynamic flow fields generated by active carpets of pusher micro-swimmers,  confined to a slick-like environment. By fixing the carpet size and varying the geometric confinement determined by hydrodynamic boundary conditions, we examine how confinement tunes the collective flow field pattern and passive transport within the fluid interface. Taking a theoretical approach and by performing numerical simulations, we show through a scale-invariant criterion that there is an optimal confinement where the active carpet induce correlated and long-ranged three-dimensional vortex-ring-like patterns with characteristic length comparable to the confinement length, enhancing passive transport to the carpet edges in a process of advective, superdiffusive process. Our findings demonstrate that the interplay between microbial aggregates and slick size heavily influences the collective flow, triggering an adaptation mechanism that might benefit both nutrient uptake survival in challenging confined viscous interfaces. 

\begin{figure}
    \centering
\includegraphics[width=1.0\linewidth]{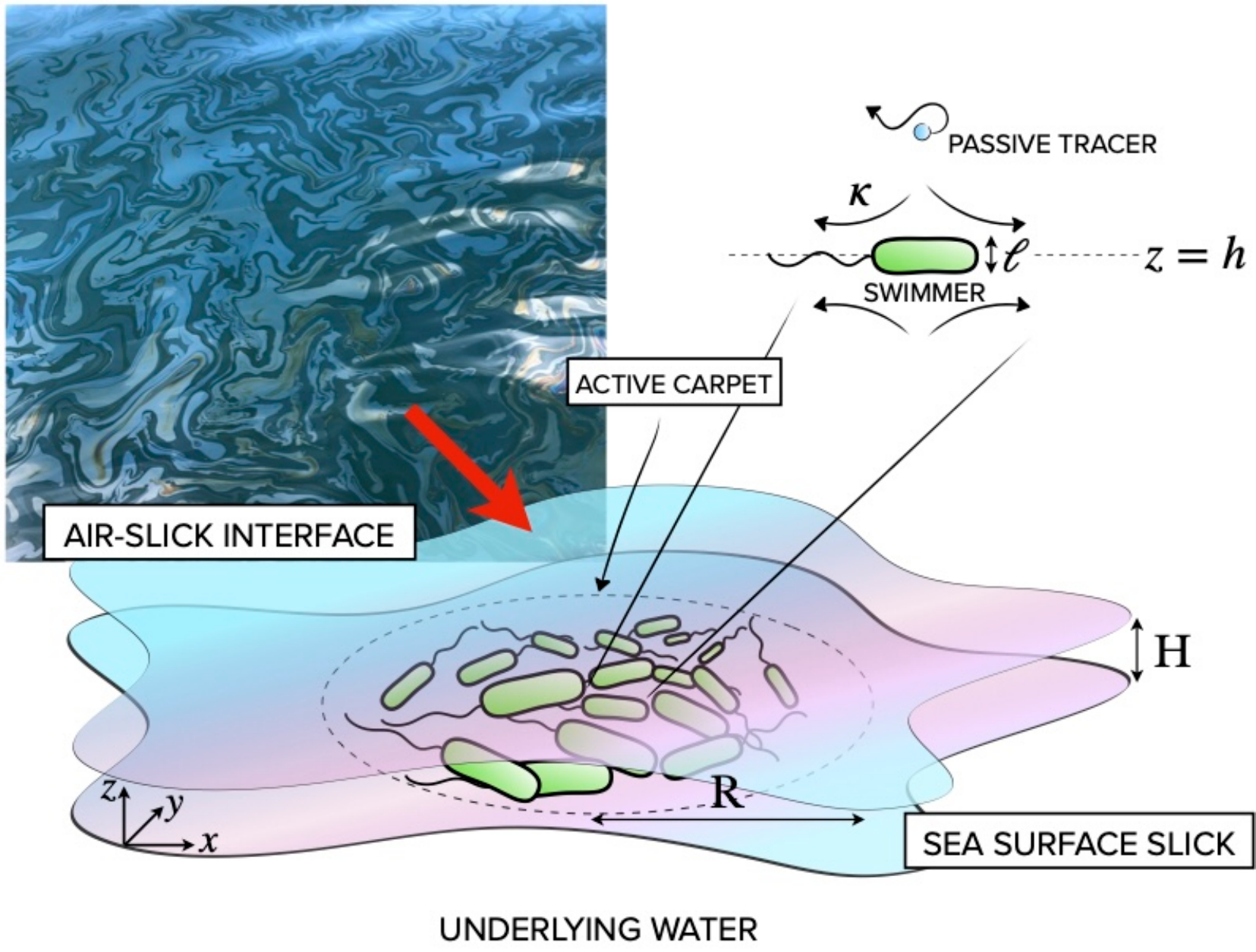}
\vspace{-0.28in}
    \caption{ {\bf  An active carpet} of size $R$ is immersed in a viscous sea slick of thickness $\mathrm{H}$, bounded above by the air-slick interface and below by the slick-water interface. The relative viscosity $\lambda$ is the ratio between the viscosities of the underlying water and the slick. The active carpet is composed of micro-swimmers of characteristic size $\ell$, suspended at a distance $h$ from the slick-water interface. Each micro-swimmer generates an extensile flow of strength $\kappa$. Background shows an image of a sea-surface slick. Credits: NASA/JPL-Caltech \cite{NASA}.}
    \label{fig:1}
\end{figure}

\vspace{-0.2in}
\section{Methods} \label{methods}

\noindent {\bf {A. Single microswimmer hydrodynamics}}\\
\noindent {\em 1. Stokelet in an unbounded fluid}

Consider a point-force $\boldsymbol{F}(\boldsymbol{x},\boldsymbol{x}_s) = \boldsymbol{f}\delta^{(3)}(\boldsymbol{x}-\boldsymbol{x}_s)$ located at $\boldsymbol{x}_s=(x_s,y_s,z_s)$ acting at a position $\boldsymbol{x}=(x,y,z)$ in a Newtonian, incompressible, unbounded viscous fluid of viscosity $\mu$. The equations for the induced flow are:
\begin{align}
    -\nabla P(\boldsymbol{x})+ \mu &\nabla^2 \boldsymbol{u}(\boldsymbol{x})+ \boldsymbol{f} \delta^{(3)}(\boldsymbol{x}-\boldsymbol{x}_s) &= \boldsymbol{0}  \\
&\nabla \cdot \boldsymbol{u}(\boldsymbol{x}) &=0
\end{align}
where $\boldsymbol{u}(\boldsymbol{x})$ is the velocity field and $P(\boldsymbol{x})$ is the pressure field. In the far-field, the fundamental singular solution for the flow is given in terms of the Oseen tensor or Stokeslet:
\begin{equation}
\mathcal{G}(\boldsymbol{x}-\boldsymbol{x}_s) = \dfrac{1}{8 \pi \mu} \left( \dfrac{\boldsymbol{\mathrm{I}}}{|\boldsymbol{x}-\boldsymbol{x}_s|} + \dfrac{(\boldsymbol{x}-\boldsymbol{x}_s)(\boldsymbol{x}-\boldsymbol{x}_s)}{|\boldsymbol{x}-\boldsymbol{x}_s|^3} \right)
\end{equation}
where $\boldsymbol{I}$ is the identity matrix. Then, the obtained velocity field is $\boldsymbol{u}^{\mathcal{S}}(\boldsymbol{x},\boldsymbol{x}_s) = \mathcal{G}(\boldsymbol{x}-\boldsymbol{x}_s) \cdot \boldsymbol{f}$.

\noindent {\em 2. Stokeslet in a floating biofilm}

A floating biofilm is a liquid layer confined between a liquid-liquid interface (L-L), and an air-liquid interface (A-L), both assumed to be non-deformable. We label this liquid layer as fluid 1, and the semi-infinite fluid beneath the liquid-liquid interface as fluid 2. For a point-force $\boldsymbol{F}(\boldsymbol{x},\boldsymbol{x}_s) = \boldsymbol{f}\delta^{(3)}(\boldsymbol{x}-\boldsymbol{x}_s)$ at a position $\boldsymbol{x}_s$, acting on the floating biofilm at a position $\boldsymbol{x}$, the equations for fluid 1 and fluid 2 are:
\begin{align}
  -\nabla P^{(1)}(\boldsymbol{x})+ \mu_1 \nabla^2 \boldsymbol{u}^{(1)}(\boldsymbol{x})+ \boldsymbol{f} \delta^{(3)}(\boldsymbol{x}-\boldsymbol{x}_s) &= \boldsymbol{0},  \\
    \nabla \cdot \boldsymbol{u}^{(1)}(\boldsymbol{x}) &=0, 
    \end{align}
    \begin{align}
    -\nabla P^{(2)}(\boldsymbol{x})+ \mu_2 \nabla^2 \boldsymbol{u}^{(2)}(\boldsymbol{x}) &= \boldsymbol{0},  \\
    \nabla \cdot \boldsymbol{u}^{(2)}(\boldsymbol{x}) &=0,
\end{align}
where $\mu_i$, $P^{(i)}(\boldsymbol{x})$,and $\boldsymbol{u}^{(i)}(\boldsymbol{x})$ are the dynamic viscosity, pressure field and fluid velocity field in each respective fluid, with $i=1,2$. The velocity fields $\boldsymbol{u}^{(1)}(\boldsymbol{x})$ and $\boldsymbol{u}^{(2)}(\boldsymbol{x})$ must satisfy appropriate boundary conditions at both limits of the floating biofilm. On the bottom, at the liquid-liquid interface, $z=0$, velocity and shear stress must be continuous, whereas on the top, at the air-liquid interface, $z=H$, the normal velocity and shear stress must vanish:
\begin{align}
\boldsymbol{u}^{(1)} &= \boldsymbol{u}^{(2)},\quad \text{at} \quad\, z=0,  \\
\mu_1 (\dfrac{\partial u_\alpha^{(1)}}{\partial z} + \dfrac{\partial u_z^{(1)}}{\partial \alpha} )  &= \mu_2 ( \dfrac{\partial u_\alpha^{(2)}}{\partial z} + \dfrac{\partial u_z^{(2)}}{\partial \alpha} ),\quad \text{at} \quad z=0,\\
    u_z^{(1)} &= 0 ,\quad \text{at} \quad\, z =H,  \\
    \frac{\partial u_\alpha^{(1)}}{\partial z} +  \frac{\partial u_z^{(1)}}{\partial \alpha} &= 0,\quad \text{at} \quad\, z=H.
\end{align}
Here $\alpha =x,y$, and $\mathrm{H}$ is the film height.

This set of equations can be solved by using the method of images, which introduces a set of appropriate image velocity fields that added to the one of an unbounded Stokeslet flow in the film generate an the effective velocity field that properly satisfy the boundary conditions. The fluid velocity field in the floating biofilm will then be the superposition of the unbounded Stokelet velocity and a film image system velocity $\boldsymbol{u}^{\text{film}}(\boldsymbol{x})$.

Here we consider $\boldsymbol{u}^{\text{film}}(\boldsymbol{x})$ to account only for the first two set of images closer to the film as they contribute the most to the hydrodynamics. 
 Then, the first image system located at a position $\boldsymbol{x}_s^{*} = (x_s,y_s,-z_s)$, accounting for the liquid-liquid interface, is given by
\begin{align}
   &\mathcal{G}^{\text{L-L}}(\boldsymbol{x},\boldsymbol{x}_s,\boldsymbol{x}_s^{*};\lambda) = - \boldsymbol{\mathrm{M}}^{\lambda} \cdot \mathcal{G}(\boldsymbol{x}-\boldsymbol{x}_s^{*}) \nonumber \\
   &+  2 \zeta_1 z_s \nabla_s \hat{e}_z \cdot \mathcal{G}(\boldsymbol{x}-\boldsymbol{x}_s^{*}) +  \zeta_1 z_s^2 \boldsymbol{ \mathrm{M}} \cdot \mathcal{G}(\boldsymbol{x}-\boldsymbol{x}_s^{*})
\end{align}
where $\lambda=\mu_2/\mu_1$ is the viscosity ratio, $\zeta_1 = \lambda/(\lambda+1)$, $\boldsymbol{\mathrm{M}}^{\lambda} = \text{diag}(\zeta_2,\zeta_2,1)$, with $\zeta_2= (\lambda+1)/(\lambda-1)$ and $\boldsymbol{\mathrm{M}}=\text{diag}(1,1,-1)$.  On the other hand, the second image system, located at a position $\boldsymbol{x}_s^{**} = (x_s,y_s,2\mathrm{H}-z_s)$, accounting for the air-liquid interface is given by
\begin{equation}
   \mathcal{G}^{\text{A-L}}(\boldsymbol{x},\boldsymbol{x}_s,\boldsymbol{x}_s^{**}(H)) =  \boldsymbol{\mathrm{M}} \cdot \mathcal{G}(\boldsymbol{x}-\boldsymbol{x}_s^{**}).
\end{equation}
The Green's function for the film is then the superposition:
\begin{align}
    \mathcal{G}^{\text{film}}(\boldsymbol{x},\boldsymbol{x}_s,\boldsymbol{x}_s^{*},\boldsymbol{x}_s^{**},\lambda) \approx &\mathcal{G}(\boldsymbol{x},\boldsymbol{x}_s)+ \mathcal{G}^{\text{L-L}}(\boldsymbol{x},\boldsymbol{x}_s,\boldsymbol{x}_s^{*},\lambda)  \nonumber \\
    + &\mathcal{G}^{\text{A-L}}(\boldsymbol{x},\boldsymbol{x}_s,\boldsymbol{x}_s^{**}(H)).
\end{align}
Thus, the first-order approximate effective velocity field induced in the floating biofilm is
\begin{equation}
    \boldsymbol{u}^{(1)}(\boldsymbol{x},\boldsymbol{x}_s,\boldsymbol{x}_s^{*},\boldsymbol{x}_s^{**};\lambda) \approx \mathcal{G}^{\text{film}}(\boldsymbol{x},\boldsymbol{x}_s,\boldsymbol{x}_s^{*},\boldsymbol{x}_s^{**}(H);\lambda) \cdot \boldsymbol{f}. 
\end{equation}

\noindent {\em {3. Stresslet in a floating biofilm}}

Swimming microorganisms are neutrally buoyant, so there are not net forces exerted over their body. Instead, a micro-swimmer of size $\ell$, swimming with speed $v_s$ and swimming orientation $\boldsymbol{p}_s$, generate thrust forces $F_0 \boldsymbol{p}_s $ that are balanced by viscous drag forces  $-F_0 \boldsymbol{p}_s$ as it translates through the medium, which can be exerted from the body or the flagella, depending on the propulsion mechanism, resulting in dipolar forces with strength $\kappa$, of extensile ($\kappa>0$) or contractile ($\kappa<0$) mechanics \cite{lauga2009hydrodynamics, saintillan2018rheology, Guzman-Lastra25}.  

Considering this, a multipole expansion performed over the generated flow across the whole body allows one to describe the hydrodynamics of the far-field dipolar velocity field, commonly known as Stresslet \cite{pushkin2013fluid}. Omitting the dependence on the force image positions in the film to emphasize the relevant parameters, the velocity field can be written as
\begin{equation}
    \dfrac{\boldsymbol{u}(\boldsymbol{x},\boldsymbol{x_s},\boldsymbol{p}_s;\lambda,H)}{8\pi \mu_1} = \kappa (\boldsymbol{p}_s\cdot\nabla_s)\boldsymbol{u}^{(1)}(\boldsymbol{x},\boldsymbol{x_s},\boldsymbol{p}_s;\lambda,H), 
\end{equation}
where $\nabla_{\!s} \equiv \partial/\partial \boldsymbol{x}_s$, $\kappa$ is the dipole strength (scaling as $\kappa \sim \ell^{2} v_s$, with units of $\mu\mathrm{m}^{3}\,\mathrm{s}^{-1}$). We set $\ell=1\,\mu m$ and $v_s=10 \,\mu m/s$, such that $\kappa=30\,  \mu m^3/s $, which value is on the lower range of the measured strength for straight-swimming \textit{E. coli} bacteria \cite{drescher2011fluid}.  \\

\begin{figure*}
    \centering
       \includegraphics[width=1.35\columnwidth]{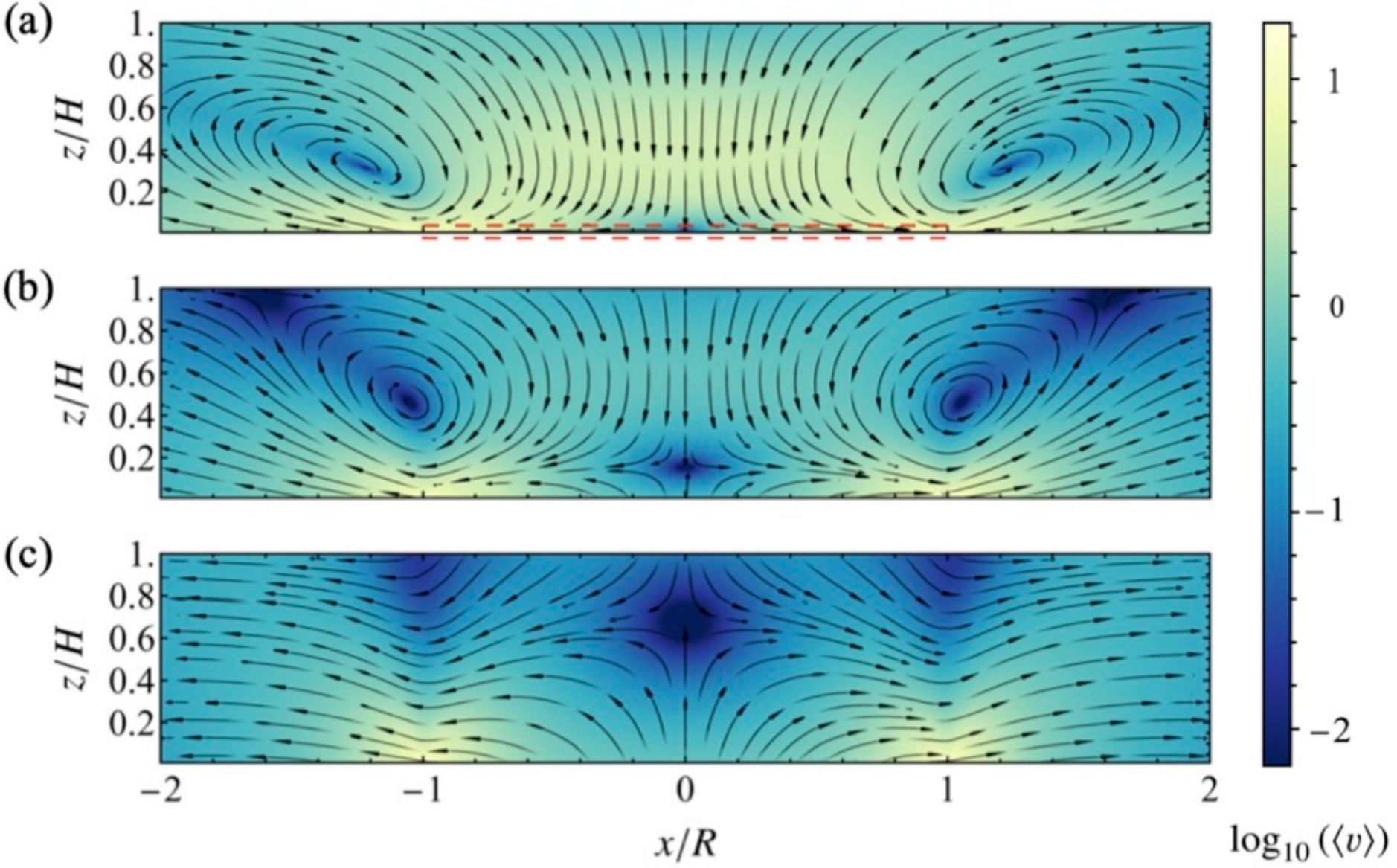}
       \vspace{-0.15in}
    \caption{\textbf{Collective average velocity field within the slick}. Panels from top to bottom correspond to a)$\mathrm{H}/R=2$ (thick slick), b)$\mathrm{H}/R=1$ and c) $\mathrm{H}/R=1/2$ (thin slick). The contours are the logarithm base 10 of the velocity field magnitude in the $y=0$ plane. The black arrows are the velocity vectors. The red dashed lines in a) indicate the diameter of the floating active carpet, $2R$.}
    \label{fig:2}
\end{figure*}

\begin{figure}
    \centering
        \includegraphics[width=0.62\columnwidth]{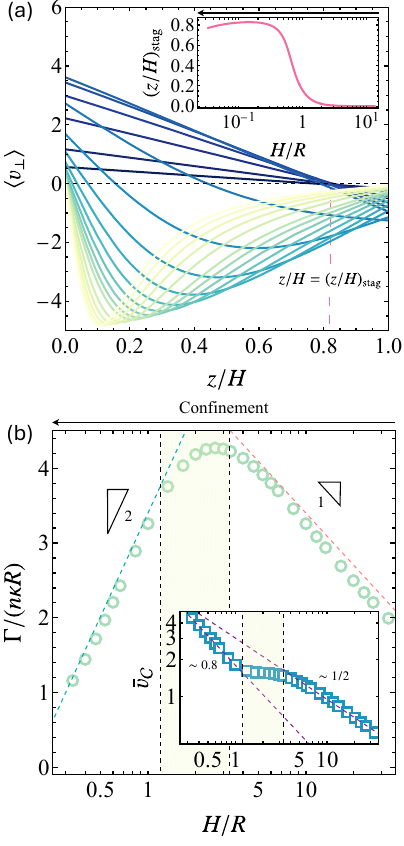}
    \vspace{-0.15in}
    \caption  { (a) \textbf{Transversal collective velocity} at the slick center for confined (blue) to non-confined cases (green). The pink dashed line is the stagnation height of the fluid velocity $(z/H)_\text{stag}$ as a reference. Inset: Analytical value of the stagnation height for a varying confinement (pink, solid line). The degree of confinement is indicated by the black arrow.
    (b) \textbf{Normalized flow circulation} over a closed curve defined as an ellipse centered at $z=H/2$ with width $(4R/3$ and height $(H-5)/2$. Inset: Average velocity over closed curve.}
    \label{fig:3}
\end{figure}


\noindent {\bf 2. Active carpet dynamics}\\
\noindent {\em 1. Definition of active carpet}

We consider a finite distribution of $N_s$ micro-swimmers moving in fluid~1, confined within a disk of radius $R$, restricted to the plane $z_s=h$, where each a position $\boldsymbol{x}_s=(\rho_s \cos \varphi_s,\rho_s \sin \varphi_s,h)$ and unit orientation $\boldsymbol{p}_s =(\cos \phi_s ,\sin \phi_s  , 0)$, and whose exerted flows are evaluated at position $\boldsymbol{x} = (\rho \cos{\varphi},\sin \varphi,z)$, with $\rho_s \in [0,R]$, and $\varphi_s,\phi_s,\varphi \in [-\pi,\pi]$. Here, we consider a constant uniform distribution $f(\boldsymbol{x}_s,\boldsymbol{p}_s) = f_s = n/2\pi$, where $n =N_s/(\pi R^2)$ is the micro-swimmer surface concentration, to which we will refer as active carpet (AC) \cite{mathijssen2018nutrient,guzman2021active,barros2025layered}. Through out this work we set the micro-swimmer density as $n \sim 0.1 \mu m^2$, which is on the higher range of the values of measured experimental $E.coli$ densities \cite{figueroa2015living,bardfalvy2024collective}. \\

\noindent {\em 2. Average velocity field and circulation}

Since the AC distribution is known, it is possible to obtain the exerted average velocity field by integrating the velocity field of a single microswimmer $\boldsymbol{u}(\boldsymbol{x}_s,\boldsymbol{x}_s,\boldsymbol{p}_s;\lambda,H)$ over it as
\begin{equation}
    \langle \boldsymbol{v}(\boldsymbol{x};\lambda,H)\rangle = \int_{\mathcal{S}}\boldsymbol{u}(\boldsymbol{x},\boldsymbol{x}_s,\boldsymbol{p}_s;\lambda,H) f_sd\boldsymbol{x}_s d \boldsymbol{p}_s.
    \label{eq:average_velocity}
\end{equation}
where $\mathcal{S}$ corresponds to the active carpet surface. The average velocity field magnitude is $\langle |\boldsymbol{v}|\rangle = \langle v\rangle $. The average circulation can be obtained directly as  $\Gamma = \oint_{\mathcal{C}} \boldsymbol{u} \cdot d\boldsymbol{\ell}$ . A simplified analytical expression for the velocity $\boldsymbol{u}(\boldsymbol{x},\boldsymbol{x}_s,\boldsymbol{p}_s;\lambda,H)$ can be found by making the approximation $h/z \ll 1$ to a first order before integration \cite{barros2025layered}. The numerical integrations shown here were performed using the Wolfram Language Mathematica \cite{Mathematica}. \\

\noindent {\em 3. Computed flows and tracer particle dynamics}

The average velocity field exerted by the AC is computed by summing over the fluid flows by all micro-swimmers:
\begin{equation}
    \langle \boldsymbol{v}(\boldsymbol{x};\lambda,H) \rangle = \sum_{i=1}^{N_s} \boldsymbol{u}_i(\boldsymbol{x},\boldsymbol{x}_{s},\boldsymbol{p}_{s};\lambda,H). 
\end{equation}

To perform this calculation, first, positions $\boldsymbol{x}_{s}(\rho_s,\varphi_s)$, and orientations $\boldsymbol{p}_{s}(\phi_s)$ are sampled by using a Smirnov transform, such that, $\rho_{s} = R \sqrt{w_{1}}$, $\varphi_{s} = -\pi +2\pi w_{2}$ and  $\phi_{s} = -\pi +2\pi w_{3}$, with $w_i \in [0,1]$ random variates drawn from standard uniform distribution, for $i \in [1,2,3]$. 

The velocity field can be characterized in all space $\boldsymbol{x}$ above the AC and below the air-liquid interface, with $h<z<H$. Hence, tracer passive particles with position $\boldsymbol{x}=\boldsymbol{x}(t)$ probe the fluid flow field, without taking into account thermal fluctuations or short-range interactions \cite{mathijssen2015tracer,guzman2021active}. 

The tracer particle's equation of motion is:
\begin{equation}
   \frac{d \boldsymbol{x}(t)}{dt} = \sum_{i=1}^{N_s}  \boldsymbol{u}_i(\boldsymbol{x}(t),\boldsymbol{x}_{s},\boldsymbol{p}_{s};\lambda,H),
\end{equation}
subject to the initial condition $\boldsymbol{x}_{\text{in}}$. To perform numerical simulations, this expression is discretized using finite differences. We approximate the solution $\boldsymbol{x}$ by the discrete-time sequence $\boldsymbol{x}_i  \approx \boldsymbol{x}(t_i)$ evaluated at the finite time $t_i = i \Delta t$, where $\Delta t$ is a sufficiently small time-step \cite{volpe2014simulation}. Finally, we solve the differential equation using an Euler scheme
\begin{equation}
    \boldsymbol{x}_i = \boldsymbol{x}_{i-1} + \sum_{i=1}^{N_s}  \boldsymbol{u}_j(\boldsymbol{x}_{i-1},\boldsymbol{x}_{s},\boldsymbol{p}_{s};\lambda,H) \Delta t.
\end{equation}
 As in Refs.~\cite{guzman2021active,aguayo2024floating,barros2025layered}, we set uniform distributions of swimmer positions $\{\boldsymbol{x}_{s}\}$ and orientations $\{\boldsymbol{p}_{s}\}$ at each iteration $i$.


\noindent {\bf C. Spatial correlation function} 

A useful observable to gain information about the flow structure is the equal-time spatial correlation function \cite{mathijssen2018nutrient,barros2025layered,fan2021effects}. We compute the (normalized) correlation for the AC average flow decomposed into its longitudinal ($\alpha = \parallel$) and transversal  ($\alpha = \perp$) components:
\begin{equation}
g_\alpha(\Delta r) = \dfrac{\langle \boldsymbol{v}_\alpha(\boldsymbol{x}) \cdot \boldsymbol{v}_\alpha(\boldsymbol{x}+\Delta r) \rangle}{ \langle \boldsymbol{v}_\alpha(\boldsymbol{x}) \cdot \boldsymbol{v}_\alpha(\boldsymbol{x}+\Delta r_0) \rangle} 
\label{eq:corr}
\end{equation}
where $\boldsymbol{v}_\parallel = v_x(\boldsymbol{x}) \hat{\boldsymbol{e}}_x + v_y(\boldsymbol{x}) \hat{\boldsymbol{e}}_y$,  $\boldsymbol{v}_\perp = v_z \hat{\boldsymbol{e}}_z$, and $\Delta r =\sqrt{\Delta x^2 + \Delta y^2 }$. The normalization is computed for the fixed separation $\Delta r_0 = h$. \\


\noindent {\bf D. Tracer's mean square displacement} 

As mentioned above, tracer particles in the slick can represent organic material, sediment, or nutrients, and tracking their trajectories over time allows us to describe the nature of the flow injected by the active carpet e.g. of advective or diffusive type. Therefore, we solve the dynamics of tracer particles and compute their MSD at different confinement levels (or slick sizes, $\mathrm{H}$)  in both directions $\alpha$:
\begin{equation}
    \text{MSD}_\alpha(t) =  \langle |\boldsymbol{x}_\alpha(t)-\boldsymbol{x}_{\alpha,0}|^2 \rangle.
\end{equation}

\begin{figure*}
    \centering
    \includegraphics[width=2.0\columnwidth]{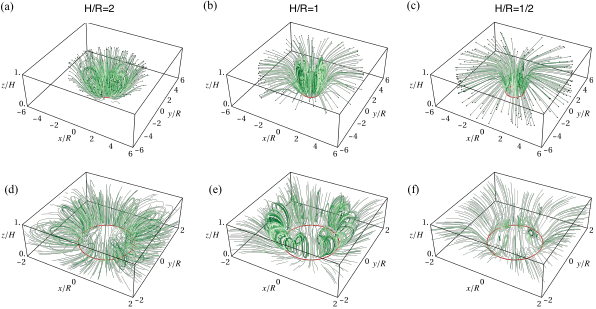}
    \vspace{-0.2in}
    \caption{\textbf{Lagrangian trajectories of tracer particles.} Top panels (a-c) show representative tracer particle trajectories for $\mathrm{H}/R=2$, $\mathrm{H}/R=1$ and $\mathrm{H}/R=1/2$ with their respective insets zooming between $[-2R,2R]$ shown in (d-f). Red circles correspond to the active carpet edge.}
    \label{fig:4}
\end{figure*}
\begin{figure*}
    \centering
     \includegraphics[width=2.0\columnwidth]{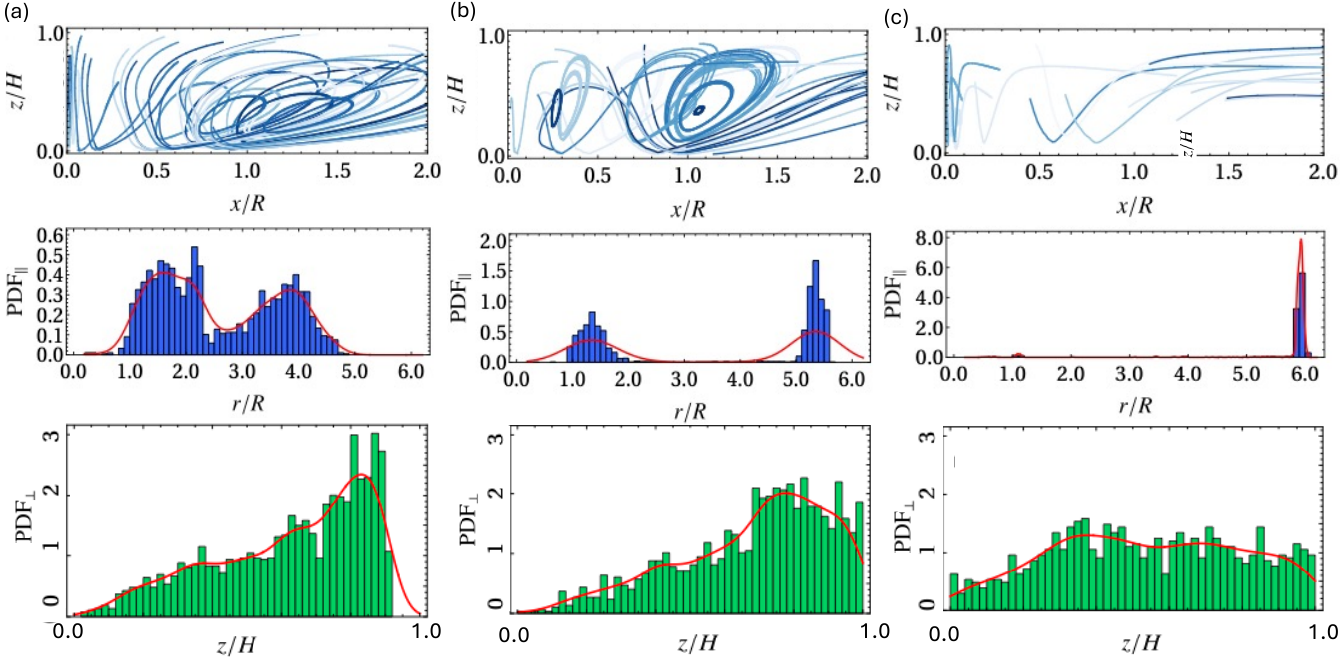}
    \vspace{-0.2in}
    \caption{ \textbf{Analysis of tracer trajectories} Top panels (a-c) show the $xz-\text{plane}$ projections of the trajectories for the cases $\mathrm{H}/R=2$, $\mathrm{H}/R=1$ and $\mathrm{H}/R=1/2$ in Fig. \ref{fig:5}. Middle panels show the radial distribution of the final positions with histograms in blue and PDF's in red solid lines for (a-c). Bottom panels show the vertical distribution of the final positions with histograms in green and PDF's in red solid lines for (a-c).}
    \label{fig:5}
\end{figure*}

\vspace{-0.2in}
\section{Results and Discussion}
We compute the average velocity field by numerically integrating Eq.\eqref{eq:average_velocity} for an AC size of $R=150 \, \mu m$ and three slick thickness: $H=300 \, \mu m$, $H= 150\, \mu m$ and $H=75\, \mu m$. These cases are analyzed in terms of the ratio $\mathrm{H}/R$, corresponding $\mathrm{H}/R =2$ (thick slick), $\mathrm{H}/R =1$ and $\mathrm{H}/R =1/2$ (thin slick), respectively. The viscosity ratio is fixed as $\lambda=1/2$, assuming that the AC is located within the highly viscous fluid. The results are shown in Fig.\ref{fig:2}.

For further analysis, we separate the velocity field $\langle v_\alpha \rangle$ into its longitudinal ($\alpha=\parallel$) and transversal ($\alpha=\perp$) components. Using the theoretical approach proposed in \cite{mathijssen2018nutrient}, we obtain an analytical expression for $\langle v_\perp  \rangle$.\\


\noindent {\bf A. Flow descriptions}

We observe that the topology of the average flow field generated by an active carpet, in a viscous film suspended between two fluid interfaces, of radius $R$ depends strongly on the geometrical confinement ratio $\mathrm{H}/R$.

In the limit of weak confinement ($\mathrm{H}/R \gg 1$), the flow exhibits a vertical drift directed towards the active carpet, accompanied by a weak recirculating component that decays with longitudinal distance (see Fig.~\ref{fig:2}(a)). This behavior was previously discussed by Mathijssen et al.~\cite{mathijssen2018nutrient} for an active carpet near a no-slip boundary in an unbounded fluid.

When the AC size becomes comparable to the slick thickness ($\mathrm{H}/R \sim 1$), the mean flow develops a vortex-ring–like pattern characterized by three stagnation points (marked in dark blue in Fig.~\ref{fig:2}(b)).

Under strong confinement ($\mathrm{H}/R \ll 1$), the flow adopts a hyperbolic topology, with a prominent central stagnation point whose vertical position shifts upward as confinement increases (see Fig.~\ref{fig:2}(c)). We therefore hypothesize that a finite-size active carpet located at the bottom interface of a fluid layer generates an hyperbolic flow whose center migrates upward as the film height $\mathrm{H}$ decreases. When $\mathrm{H}$ is large, the stagnation center lies near the lower boundary; as $\mathrm{H}$ approaches $R$, it shifts toward the upper interface, giving rise to a vortex-ring transition.

To analyze the transition between these flow regimes, we computed the vertical displacement of the central stagnation point of the hyperbolic flow as a function of confinement. Specifically, we evaluated the average vertical velocity $v_{\perp}$ at the active carpet center ($\rho = 0$) for different slick thicknesses $\mathrm{H}$.

Fig.~\ref{fig:3}a shows that $v_{\perp}$ takes both positive and negative values depending on $\mathrm{H}/R$, revealing three distinct regimes. For strong confinement ($\mathrm{H}/R \le 2/3$), the flow remains hyperbolic, with the stagnation point located in the upper middle region, where the vertical velocity is positive throughout most of the film (blue lines in the figure). As $\mathrm{H}/R$ increases, the stagnation point moves downward toward the mid-lower region, marking the onset of the vortex-ring regime, where $v_{\perp}$ changes sign smoothly (light-blue region). Finally, for weak confinement, the stagnation point approaches the lower interface near the active carpet, where a strong downward drift develops, indicated by a pronounced negative minimum in $v_{\perp}$ (yellow and green lines).

The inset in Fig.~\ref{fig:3}a presents the analytical position of the stagnation point at $\rho = 0$ as a function of $\mathrm{H}/R$, showing a continuous transition from values near the upper interface to those near the lower boundary.

Next, we evaluated the circulation dynamics around the point $(\rho = R,\, z = H/2)$. We defined an ellipsoidal contour of horizontal radius $5R/3$ and vertical semi-axis $(H - 5)/2$, along which we computed the circulation $\Gamma$ and the mean tangential velocity $\bar{v}_C$. The results, shown in Fig.~\ref{fig:3}b, are normalized by $n \kappa R$, representing the flux per unit line segment generated by a density $n = N_s/\pi R^2$ of micro-swimmers exerting dipolar flows of strength $\kappa$ within a disk of radius $R$.

The circulation exhibits a non-monotonic dependence on confinement: it increases approximately quadratically with $\mathrm{H}/R$, reaches a maximum for $1 < H/R < 3$, and subsequently decays roughly linearly with $\mathrm{H}/R$.

Fig.~\ref{fig:3}b inset shows that the average circulation velocity $\bar{v}_C$ spans the three identified flow regimes:
\begin{itemize}
\item $\bar{v}_C \propto (H/R)^{-0.8}$ for the hyperbolic flow,
\item $\bar{v}_C \approx 1.5 \mu m/s$ for the vortex-ring regime,
\item $\bar{v}_C \propto (H/R)^{-1.2}$ for the drift-dominated flow.
\end{itemize} 

\begin{figure}
    \centering
    \includegraphics[width=1\linewidth]{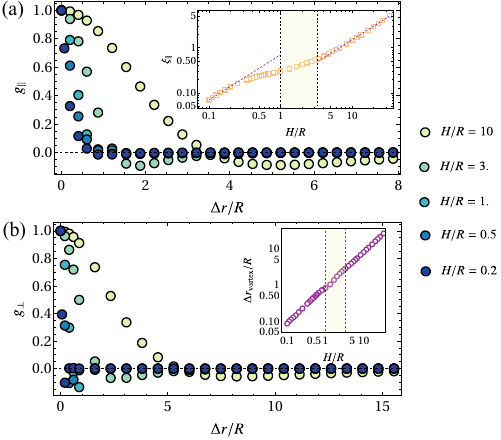}
        \vspace{-0.3in}
    \caption{\textbf{Mean squared displacement of tracer particle trajectories for a varying confinement.} (a) Longitudinal and (b) transversal MSD as a function of time for the characteristic cases $\mathrm{H}/R=2$ (green circles), $\mathrm{H}/R=1$ (blue squares) and $\mathrm{H}/R=1/2$ (sky-blue triangles). Markers correspond to results from numerical simulations. The white triangle shows a power-law $t^\alpha$ with $\alpha=2$.}
    \label{fig:6}
\end{figure}
\begin{figure}
    \centering
    \includegraphics[width=0.65\linewidth]{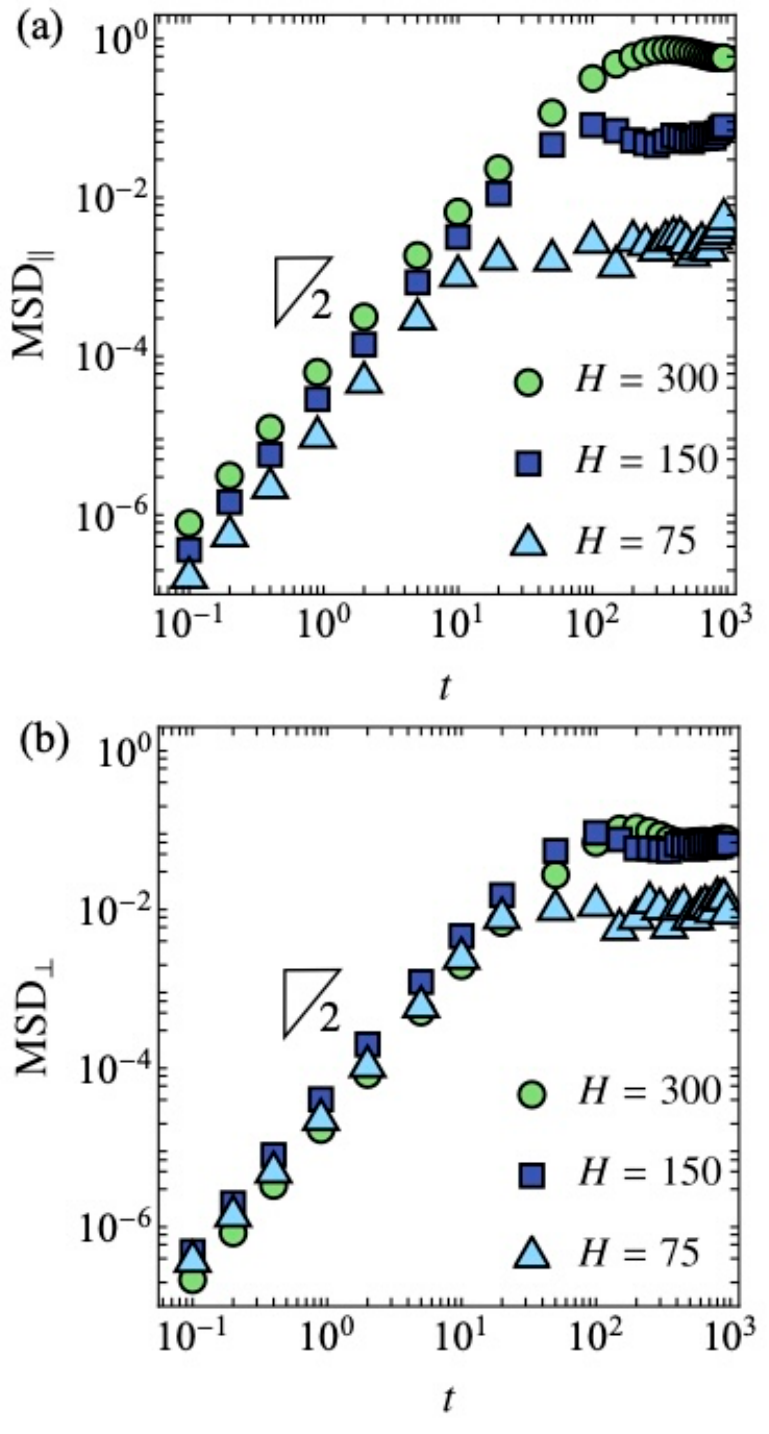}
    \vspace{-0.25in}
    \caption{\textbf{Equal-time normalized spatial correlation function of the active carpet flow for different confinements}. Correlation of the (a) radial (Inset: Correlation length for several confinement values. The vortex ring lives in the yellow region. The dashed purple lines are power laws $(H/R)^{\alpha}$ with exponent $\alpha=1$.) and (b) vertical (Inset: Normalized distance to the minimum correlation value as a function of the parameter $\mathrm{H}/R$) velocities $\boldsymbol{u}_\parallel$, $\boldsymbol{u}_\perp$, respectively, as function of the normalized distance $\Delta r/R$.}
    \label{fig:7}
\end{figure}


\noindent {\bf B. Particle distributions}

To investigate the transport generated in different flow regimes, we analyzed the resulting particle distributions. We first examined the Lagrangian trajectories of passive tracer particles uniformly distributed within a simulation box of size $\Delta V = 3R \times 3R \times (H - 3)$. Fig.~\ref{fig:4} shows the trajectories followed by the particles in the three regimes corresponding to $\mathrm{H}/R = 2$, $1$, and $0.5$, respectively.

For weak confinement, tracers are advected along an elongated vortex ring that induces a strong drift near the center of the active carpet. This suggests that, if the tracer particles represent nutrients or essential molecules for the microorganisms, such flow enhances nutrient acquisition at the center of the active carpet, thereby supporting microorganism survival in this region.


When the confinement is comparable to the active carpet size ($\mathrm{H}/R \sim 1$), particles follow the vortex ring streamlines, resulting in a depleted central region and an accumulation near the carpet edges. This is consistent with the experimental observations reported in Ref.~\cite{hokmabad2025spatial} where microorganisms located at the cluster center exhibited anoxic conditions, while those at the periphery are active due to nutrient availability.

In the case of strong confinement, the particles are expelled from the vicinity of the active carpet. We can thus infer that in this regime, nutrient uptake by the carpet is reduced, which may explain why such configurations are rarely observed in natural environments, as also suggested by Ref.~\cite{wei2025confinement}.

Additionally, Fig.~\ref{fig:5} presents the final particle position distributions in both the horizontal (blue) and vertical (green) directions for the three confinement regimes. Vertically, the particle distribution is nearly uniform under strong confinement, while for intermediate and weak confinement, a weak peak emerges near the upper fluid interface. A close-up of the particle trajectories in the $x$–$z$ plane is also provided for clarity.

To quantitatively characterize particle transport, we measured the mean squared displacement (MSD) as described in Sec. {\bf III-D} (see Fig.~\ref{fig:6}). We focused on tracer particles trapped within the long-ranged vortical flows near the edge of the active carpet, highlighted by the red circle in Fig.~\ref{fig:5}. For each confinement condition, we computed the MSD of the particles corresponding to the first peak of the longitudinal distribution $\text{PDF}_\perp$. 

In the longitudinal plane, $\text{MSD}_\parallel$, the particles exhibit a ballistic regime at very short times, which rapidly transitions to a steady oscillatory behavior. This indicates that the tracers are confined to periodic displacements along the vortex-ring structures near the carpet boundary.

For the vertical component, $\text{MSD}_\perp$, we observe a distinct behavior: particles initially move ballistically with a velocity nearly independent of the confinement $\mathrm{H}$, presumably scaling with the vertical flux generated by the active carpet, $v_z \propto n \kappa$~\cite{mathijssen2018nutrient}. At longer times, the MSD saturates to a plateau with value determined by the vertical confinement height $\mathrm{H}$.

Overall, in the vicinity of the active carpet and in both directions, particle transport is dominated by advective processes governed by the flow topology generated by the active carpet, consistent with the behaviors we described in the previous subsection.

Following Sec. {\bf III-C} and using the analytical expression from Eq.~\eqref{eq:corr}, we calculated the pair correlation functions in both the parallel and vertical directions for two tracer particles separated by different radial distances $\Delta r / R$. Fig.~\ref{fig:7} presents the results for various confinement ratios $\mathrm{H}/R$, with yellow tones indicating weak confinement and dark blue representing strong confinement.

The insets show the normalized correlation length $\xi_j / R$, obtained from a Gaussian fit to the velocity pair correlations of the form $g_\parallel (\Delta r/R) \sim e^{-((\Delta r/R)^2/2\xi_\parallel^2)}$. This length scale represents the characteristic size of the coherent flow structures, which increases with the slick thickness $\mathrm{H}$. In particular, for the parallel fluxes, we observe the formation of small vortices at low $\mathrm{H}$ and larger swirls as $\mathrm{H}$ increases. The yellow region highlights the range associated with ring-like structures. In the vertical direction, by contrast, the correlation length increases approximately linearly with the confinement height $\mathrm{H}$.

\section{Conclusions}
We have investigated how vertical confinement controls the flows generated by a finite active carpet of flagellated micro-swimmers, modeled as force dipoles, located near the lower interface of a viscous surface slick of thickness $\mathrm{H}$ and viscosity ratio $\lambda$. Fixing $\lambda$ and varying the geometrical confinement $\mathrm{H}/R$ (with $R$ the carpet radius), our analytical approximations and numerical simulations reveal that coherent flow structures emerge purely from confinement and from the hydrodynamic distortions induced by the swimmers.

For strong confinement ($\mathrm{H}/R \ll 1$), the mean flow adopts a hyperbolic topology, similar to the flow generated by strongly confined \textit{Chlamydomonas reinhardtii} observed experimentally in Ref.~\cite{mondal2021strong}, with a central stagnation point at $\rho = 0$ displaced upward toward the free surface.
This configuration generates an outward jet that ejects particles from the carpet neighborhood, implying reduced nutrient uptake and providing a possible explanation for the scarcity of such configurations in natural settings~\cite{wei2025confinement}. As confinement is relaxed, the stagnation point shifts toward the mid-plane ($z_{\text{stag}} \sim H/2$), and for intermediate confinement ($\mathrm{H}/R \sim 1$) a vortex-ring-like circulation develops whose characteristic length scales with $\mathrm{H}$. This ring recirculates suspended tracers from the center toward the carpet edges, potentially enhancing lateral replenishment and edge activity~\cite{hokmabad2025spatial}. Finally, under weak confinement ($\mathrm{H}/R \gg 1$), the stagnation point moves back toward the lower interface, and the flow exhibits a strong drift directed toward the carpet, consistent with active-carpet pumping and substrate-ward currents that promote replenishment~\cite{mathijssen2018nutrient}.

Quantitatively, we find that the vortex-ring radius increases with $\mathrm{H}$, and that the resulting edgeward recirculation modifies particle transport: vertical displacements saturate at a plateau set by $\mathrm{H}$, while longitudinal displacements remain advective near the carpet boundary. Together with recent analyses of layered microenvironments and semi-infinite films~\cite{barros2025layered}, our results indicate that finite, coherent vortex-ring-like structures and the associated transport pathways can be selected by the single geometrical control parameter $\mathrm{H}/R$. This identifies an optimal colony size for a given $\mathrm{H}$ that maximizes advective delivery to carpet edges, suggesting testable ecological implications for colony formation, spreading, and persistence in geometrically constrained habitats such as surface slicks.

Finally, we note that a finite active carpet may also serve as a minimal physical model for localized ciliary or flagellar activity on the surfaces of larger bodies or cells, where similar flow structures and confinement effects can emerge~\cite{brumley2014flagellar,wan2020reorganization}. This connection opens a broader perspective for applying active-carpet hydrodynamics to biological transport processes at both microscopic and mesoscopic scales.

{\bf ACKNOWLEDGEMENTS}  Support is acknowledged from the ANID-Fondecyt grant no. 1250913 (F.A.B., I.S. F.G.-L.) and the Simons Foundation grant no. 639018 (E.L.).

\bibliography{bibliography}

\end{document}